\documentclass{moriond}

\usepackage{amsmath}
\usepackage{xspace}

\def\Journal#1#2#3#4{{#1} {\bf #2}, #3 (#4)}

\def\NPB{{\em Nucl. Phys.} B}
\def\PLB{{\em Phys. Lett.}  B}

\def\PRD{{\em Phys. Rev.} D}

\newcommand{\pt}{\ensuremath{p_{\text{T}}}\xspace}
\newcommand{\MET}{\ensuremath{E_\text{T}^\text{miss}}\xspace} % |Missing ET|
\newcommand{\met}{\ensuremath{E_\text{T}^\text{miss}}\xspace} % |Missing ET|
\newcommand{\ttbar}{\ensuremath{t\bar{t}}\xspace}
\newcommand{\ttbarV}{\ensuremath{t\bar{t}V}\xspace}
\newcommand{\ttbarH}{\ensuremath{t\bar{t}H}\xspace}
\newcommand{\brtht}{\ensuremath{BR(T\rightarrow Ht)}\xspace}
\newcommand{\brtwb}{\ensuremath{BR(T\rightarrow Wb)}\xspace}
\newcommand{\brtzt}{\ensuremath{BR(T\rightarrow Zt)}\xspace}
\newcommand{\brbwt}{\ensuremath{BR(B\rightarrow Wt)}\xspace}
\newcommand{\brbhb}{\ensuremath{BR(B\rightarrow Hb)}\xspace}

\newcommand{\mtlep}{\ensuremath{m_T^\mathrm{lep}}\xspace}
\newcommand{\mthad}{\ensuremath{m_T^\mathrm{had}}\xspace}

\def\be{\begin{equation}}
\def\ee{\end{equation}}
\def\bea{\begin{eqnarray}}
\def\eea{\end{eqnarray}}

\newcommand{\Photo}{}

\begin{document}
\vspace*{4cm}
\title{SEARCHES FOR NEW HEAVY QUARKS IN ATLAS}
% \linenumbers

\author{ N.~NIKIFOROU, ON BEHALF OF THE ATLAS COLLABORATION}

\address{Department of Physics, University of Texas at Austin}

\maketitle\abstracts{
A search for new heavy quarks focusing on recent vector-like quark searches with the ATLAS detector 
at the CERN Large Hadron Collider is presented.
Two recent searches targeting the pair production of type vector-like quarks are described. The 
first search is sensitive to vector-like up-type quark ($T$)
decays to a $t$ quark and either a Standard Model Higgs boson or a $Z$ boson. The second 
search is primarily sensitive to $T$ decays to $W$ boson and a $b$ quark. Additionally, the results can be interpreted for alternative VLQ decays.
}

\section{Introduction}

Although appealing, a straightforward addition of a fourth generation of Standard Model (SM) 
quarks is excluded by experimental observation;
such an addition would contribute to the SM Higgs boson production via fermion loops and would not 
be 
compatible with the observed~\cite{atlasHiggs,cmsHiggs} Higgs boson production cross-section.
For this reason and others, latest searches for new heavy quarks in ATLAS~\cite{atlasPaper} have 
targeted Vector-Like 
Quarks (VLQs). VLQs are proposed in various 
models~\cite{littlestHiggs,littleHiggsReview,compHiggsScal,minimalCompHiggs} of
new physics beyond the SM. They are coloured spin-1/2 fermions of which the left- and right-handed 
components transform the same way under gauge transformations~\cite{aguila,aguilar}. 
VLQs evade limitations imposed on chiral quark extensions of the SM, however they can mix with their 
SM quark counterparts and regulate the Higgs boson mass-squared
divergence. They therefore provide an attractive mechanism to solve the hierarchy problem. 

VLQs can have charges analogous to their SM quark counterparts, such as the $T$ and $B$ VLQs with 
charge $q=2/3e$ and $-1/3e$, respectively, 
or more exotic charges, such as in the case of the $X$ and $Y$ VLQs, with charge $q=5/3e$ and 
$-4/3e$, respectively, where $e$ is the charge of the electron.
The various VLQs can be arranged in $SU(2)$ singlets or multiplets. The VLQs can decay to the $W$, 
$Z$, and Higgs ($H$) bosons with branching ratios
which depend on the model and, in general, decays to third generation SM quarks are favoured. As a 
consequence, it is usually assumed in searches that
the VLQs couple exclusively to the $t$ and $b$ quarks. 

\begin{figure}[ht]
\begin{minipage}{0.5\linewidth}
\centerline{\includegraphics[height=3cm]{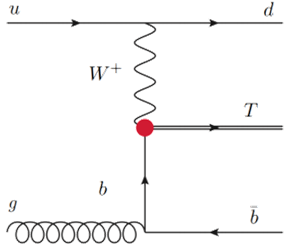}}
\end{minipage}
\hfill
\begin{minipage}{0.5\linewidth}
\centerline{\includegraphics[height=3cm]{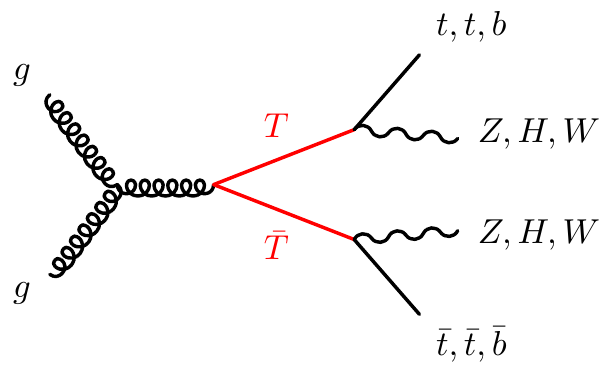}}
\end{minipage}
\caption[]{Example Feynman diagrams for single VLQ production  (left) and pair-production (right)}
\label{fig:feyndiag}
\end{figure}

The new quarks could be produced either singly or in pairs. The single production mechanism 
dominates for high VLQ masses~\cite{handbook}. For single production, the production cross-sections
depend strongly on the model, since the mechanism requires the mediation of a gauge boson, as shown on the left of 
Figure~\ref{fig:feyndiag}. 
Conversely, the pair-production mechanism, proceeding via gluon-splitting as shown on the right of
Figure~\ref{fig:feyndiag}, is generally considered model-independent
and dominates up to VLQ masses of the order of 800~GeV. As a 
consequence, the recent ATLAS searches presented in these
proceedings have focused on VLQ pair-production. Two searches are discussed in the following 
sections: a search for the pair production of up-type VLQs in events with multiple 
$b$-jets~\cite{htx} and a search for the pair production of VLQs decaying to a high-\pt $W$-boson 
and a $b$-quark in events with leptons and jets~\cite{wbx}. Both analyses are using the 
36.1~$\mathrm{fb}^{-1}$ of proton-proton collisions delivered by the LHC at $\sqrt{s} = 
13~\mathrm{TeV}$ in 2015 and 2016 and recorded by the ATLAS detector.

\section{Search for pair production of up-type vector-like
quarks and for four-top-quark events in final states
with multiple $b$-jets with the ATLAS detector ($T\bar{T}\rightarrow Ht + X$)}

The search targets $T\bar{T}$ pair production where at least one of the VL $T$ decays to a $H$ boson or 
a $Z$ boson with the boson decaying to either to a pair of $b$ quarks ($T\rightarrow H(\rightarrow 
b \bar{b})t$) or a pair of neutrinos ($T\rightarrow Z(\rightarrow \nu\bar{\nu})t$), respectively. 
Additionally, the search investigates anomalous four-top-quark production in the context of an effective field theory
model and in a universal extra dimensions model, however this analysis is not discussed in these 
proceedings.

The analysis takes advantage of the presence of multiple jets, $b$-tagged jets, $t$-tagged jets, 
$H$-tagged jets, and $\MET$. Jets are reconstructed using the anti-$k_t$ algorithm~\cite{antikt} 
with a radius parameter $R=0.4$. Jets containing $b$ hadrons are tagged using a working point with 
77\% efficiency measured in simulated \ttbar events.  
Large-radius jets are obtained by 
reclustering~\cite{reclustering} the $R=0.4$ jets using the anti-$k_t$ algorithm with a radius 
parameter $R=1.0$. By ``$t$-tagged'' and ``$H$-tagged'' jets we refer to large-radius jets 
identified with decaying $t$ quark or $H$ boson candidates by making requirements on their 
transverse momentum, mass, and number of constituents. 

The analysis is split into two channels, 
with initial preselection criteria as shown in Table~\ref{tab:htx:preselection}, taking 
advantage of the 
event characteristics of the two targeted signatures. The ``1-lepton (1L) channel'' requires 
exactly one isolated electron or muon and has a modest requirement on \MET, while the ``0-lepton 
(0L) channel'' requires the absence of any isolated electrons or muons and $\MET> 
200~\mathrm{GeV}$.

\begin{table}[ht]
 \centering
  {%
\newcommand{\mc}[3]{\multicolumn{#1}{#2}{#3}}
% \begin{center}
\begin{tabular}{l|c|c}
\hline\hline
\mc{3}{c}{Preselection requirement}\\ \hline
Requirement & 1-lepton channel & 0-lepton channel\\
Trigger & Single-lepton trigger & \met trigger\\
Leptons & $=1$ isolated $e$ or $\mu$ & $=0$ isolated $e$ or $\mu$\\
Jets & $\geq5$ jets & $\geq6$ jets\\
$b$-tagging & $\geq2$ $b$-tagged jets & $\geq2$ $b$-tagged jets\\
\met & $\met > 20~\mathrm{GeV}$ & $\met > 200~\mathrm{GeV}$\\
Other \met-related & $\met+m_\mathrm{T}^{W}$ & $\Delta\phi_\mathrm{min}^\mathrm{4j}  > 0.4$\\
\hline \hline
 \end{tabular}
%  \end{center}
}%
 \caption[]{Summary of preselection requirements for the 1-lepton and 0-lepton channels in the 
 $Ht +X$ analysis~\cite{htx}. Here 
$m_\mathrm{T}^W$  is the transverse mass of the lepton and the \met, and 
$\Delta\phi_\mathrm{min}^\mathrm{4j}$  is the minimum azimuthal separation between the 
$\vec{E}_\mathrm{T}^\mathrm{miss}$ vector and each of the four highest-\pt jets.}
 \label{tab:htx:preselection}
\end{table}

The two channels are further 
divided into search regions according to the number of $b$-, $t$-, and $H$-tagged jets in the event 
as well as the overall jet multiplicity. In addition, for the 0L  channel, the regions are 
categorized in ``high-mass'' (HM) and ``low-mass'' (LM) regions depending on whether the selected 
events satisfy (HM) or fail (LM) the requirement $m_\mathrm{T,min}^b >160~\mathrm{GeV}$, where 
$m_\mathrm{T,min}^b$ is the minimum transverse mass formed with the \met and any of the 2 (or 3) 
$b$-jets. In total, 12 (22) search regions are defined for the 1L (0L) channel.

The final discriminant is the effective mass ($m_\mathrm{eff}$), defined as the scalar sum of the 
$\pt$ of the lepton, jets, and \met present in the events. As can be 
seen in Figure~\ref{fig:htx:searchregions}, the background is dominated by $t\bar{t}$+jets events 
after preselection. Small contributions arise from single-top-quark, $W$/$Z$+jets, multijet and 
$WW$, $WZ$, $ZZ$ production, as well as from the associated production of a vector boson $V$ ($V = 
W, Z$) or a $H$ boson and a \ttbar pair (\ttbarV and \ttbarH). All backgrounds are estimated 
using samples  of simulated events and initially normalized to their theoretical cross sections, 
with the exception of the multijet background, which is estimated using data-driven methods.
{%
\begin{figure}
\begin{minipage}{0.4\linewidth}
\centerline{\includegraphics[width=0.9\linewidth]{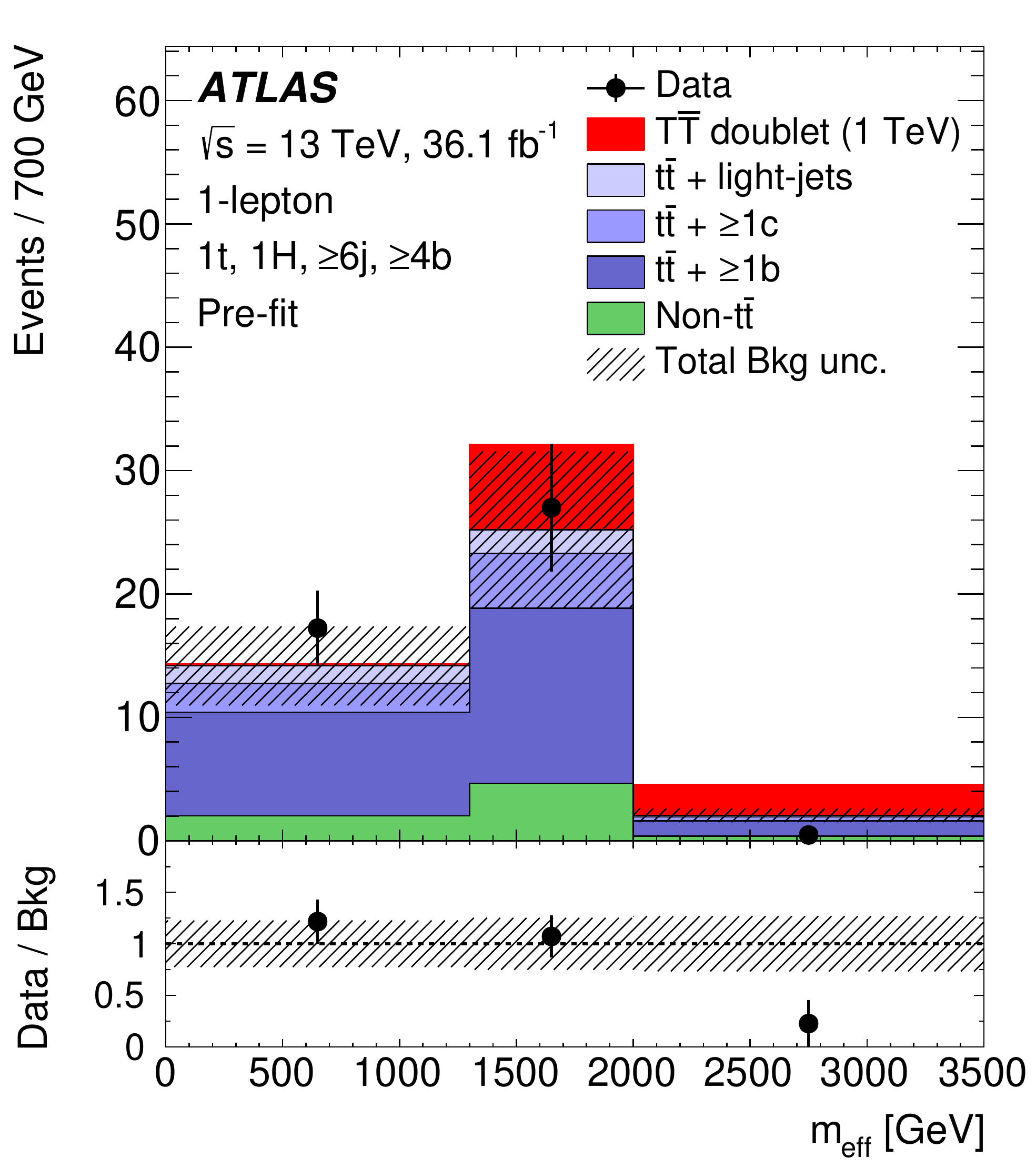}}
\end{minipage}
\hfill
\begin{minipage}{0.6\linewidth}
\centerline{\includegraphics[width=1.\linewidth]{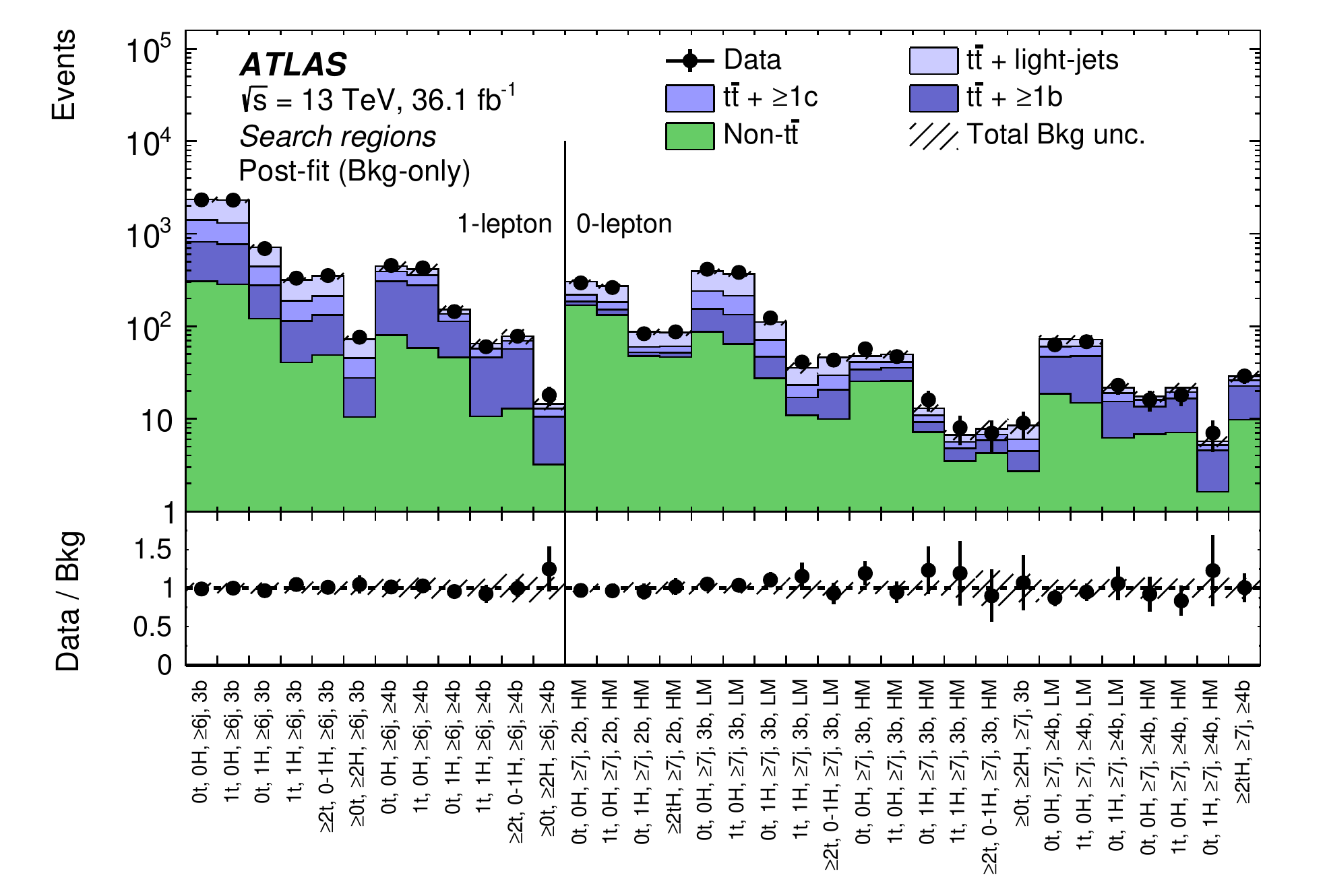}}
\end{minipage}
\caption[]{Left: Distribution of the final discriminant for data, expected background, and a signal 
model with  $m_T=1~\mathrm{TeV}$ in an example search region before the background-only fit for 
the $Ht + X$ analysis~\cite{htx}. Right: Compatibility of the background expectation with the data 
in the 
search regions after the background-only fit in the $Ht + X$ analysis~\cite{htx}.}
\label{fig:htx:searchregions}
\end{figure}
}%
A combined maximum likelihood fit of the $m_\mathrm{eff}$ distribution is performed to the data in 
all search regions to determine the normalization of the backgrounds. As can be seen in 
Figure~\ref{fig:htx:searchregions}, the background expectation has a very good agreement with the 
data in all search regions, after the background-only fit. The background modelling is further 
checked in 10 (16) additional validation regions made orthogonal by requiring exactly 5 (6) jets in 
for the 1L (0L) channel. The main systematic uncertainties, which vary by search region, are the 
modelling of the \ttbar background, flavor tagging uncertainties, and background normalization 
uncertainties.

\begin{figure}
\begin{minipage}{0.5\linewidth}
\centerline{\includegraphics[height=5cm]{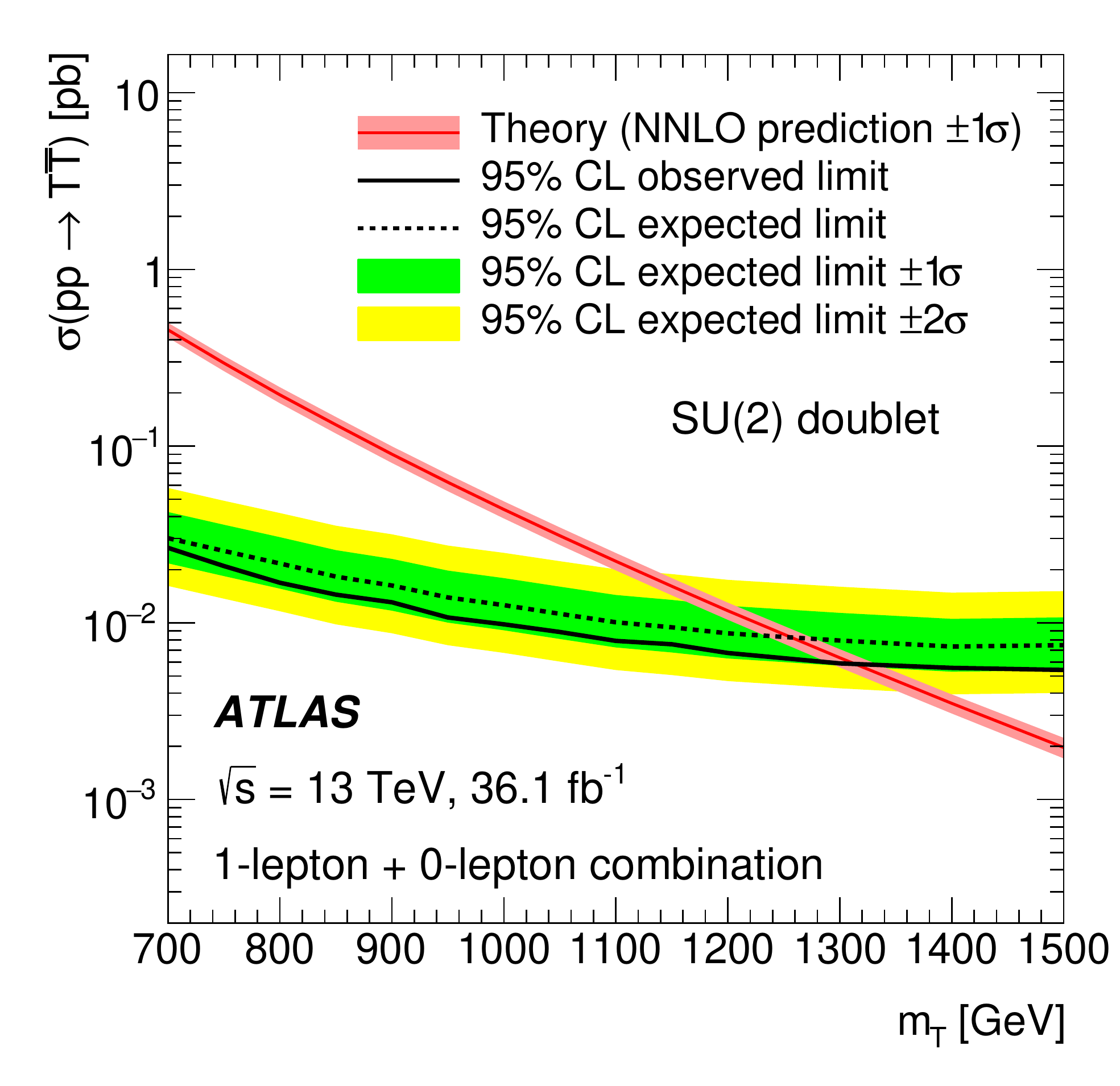}}
\end{minipage}
\hfill
\begin{minipage}{0.5\linewidth}
\centerline{\includegraphics[height=5cm]{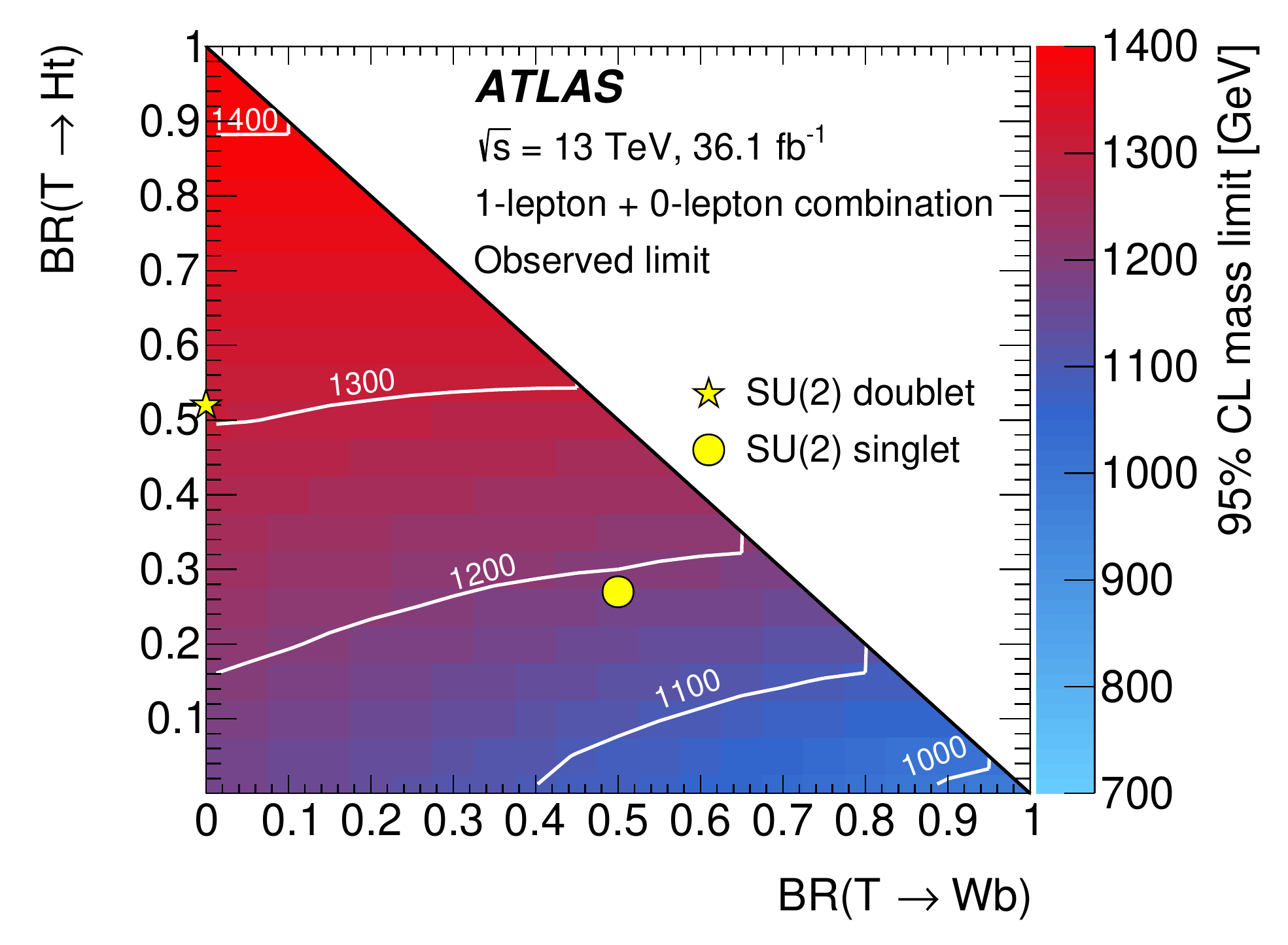}}
\end{minipage}
\caption[]{Combined expected and observed limits as a function of VL $T$ mass for an example 
BR scenario (left) and combined observed mass limits in the two-dimensional BR plane (right) for 
the $Ht+X$ analysis~\cite{htx}.}
\label{fig:htx:limits}
\end{figure}

Given the compatibility of the data and the background expectation, 95\% CL limits are set on the 
pair production of VL $T$ separately for the two channels, as well as combined. The excluded VL $T$ 
masses depend on the branching ratios assumed by the model under investigation. For example, as 
shown in Figure~\ref{fig:htx:limits},  VL $T$ 
masses up to 1.3~TeV are excluded for the SU(2) doublet model which assumes $BR(T\rightarrow 
Ht)\approx BR(T\rightarrow Zt)\approx 0.5$. For 
a model assuming a VL $T$ exclusively decaying to $Ht$ ($BR(T\rightarrow Ht)=1$) masses up to 
1.4~TeV are excluded. 

However, a more general interpretation of the results can be performed by reweighting the signal 
samples to other BR compositions to obtain two-dimensional limits on the BR plane, as shown on the 
right of Figure~\ref{fig:htx:limits}. Under the assumption $\brtht+\brtwb+\brtzt=1$, each point on 
the plane indicates a model with a given BR composition. The colour scale indicates the highest 
excluded mass at the given BR composition. As can be seen, the highest excluded masses are near 
point $(0,1)$ where $\brtht=1$ which is expected given the optimization of the search for the 
$T\rightarrow Ht$ and $Zt$ decays.

\section{Search for pair production of heavy vector-like
quarks decaying to high-\pt $W$ bosons and $b$ quarks
in the lepton-plus-jets final state in $pp$ collisions at \mbox{$\sqrt{s}=13~\mathrm{TeV}$} with 
the ATLAS detector ($Q\bar{Q}\rightarrow Wb+X$)}

The second search primarily targets $T\bar{T}$ production where at least one of the VL 
$T$ decays 
via $T\rightarrow Wb$. Events are initially required to have exactly one lepton (either an electron 
or a muon), to have at least three jets of which at least one is required to be $b$-tagged, and to 
have $\met>60~\mathrm{GeV}$. The events are further required to have at least one high-\pt $W$ 
candidate 
which decays hadronically, labelled $W_\mathrm{had}$. The selection is therefore optimized for the 
decay $T\bar{T}\rightarrow W_\mathrm{had} W_\mathrm{lep}bb$, where $W_\mathrm{lep}$ denotes a $W$ 
decaying leptonically. The four-momentum of the neutrino is analytically determined using  the 
missing transverse momentum vector $\vec{E}_\mathrm{T}^\mathrm{miss}$ and constraints of the 
lepton-neutrino system from the mass of the $W$ boson.

\begin{figure}
\begin{minipage}{0.5\linewidth}
\centerline{\includegraphics[width=0.7\linewidth]{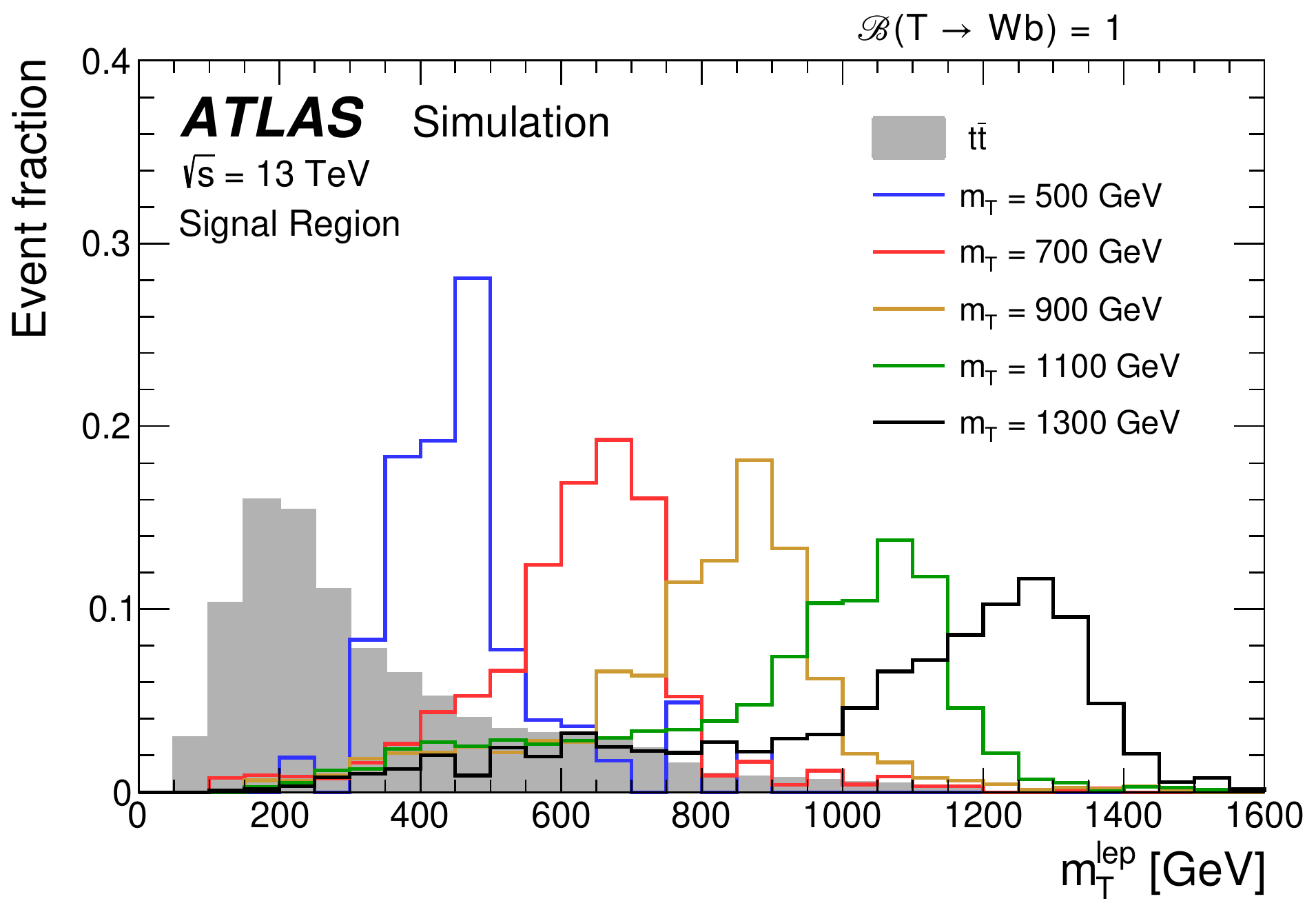}}
\end{minipage}
\hfill
\begin{minipage}{0.5\linewidth}
\centerline{\includegraphics[width=0.7\linewidth]{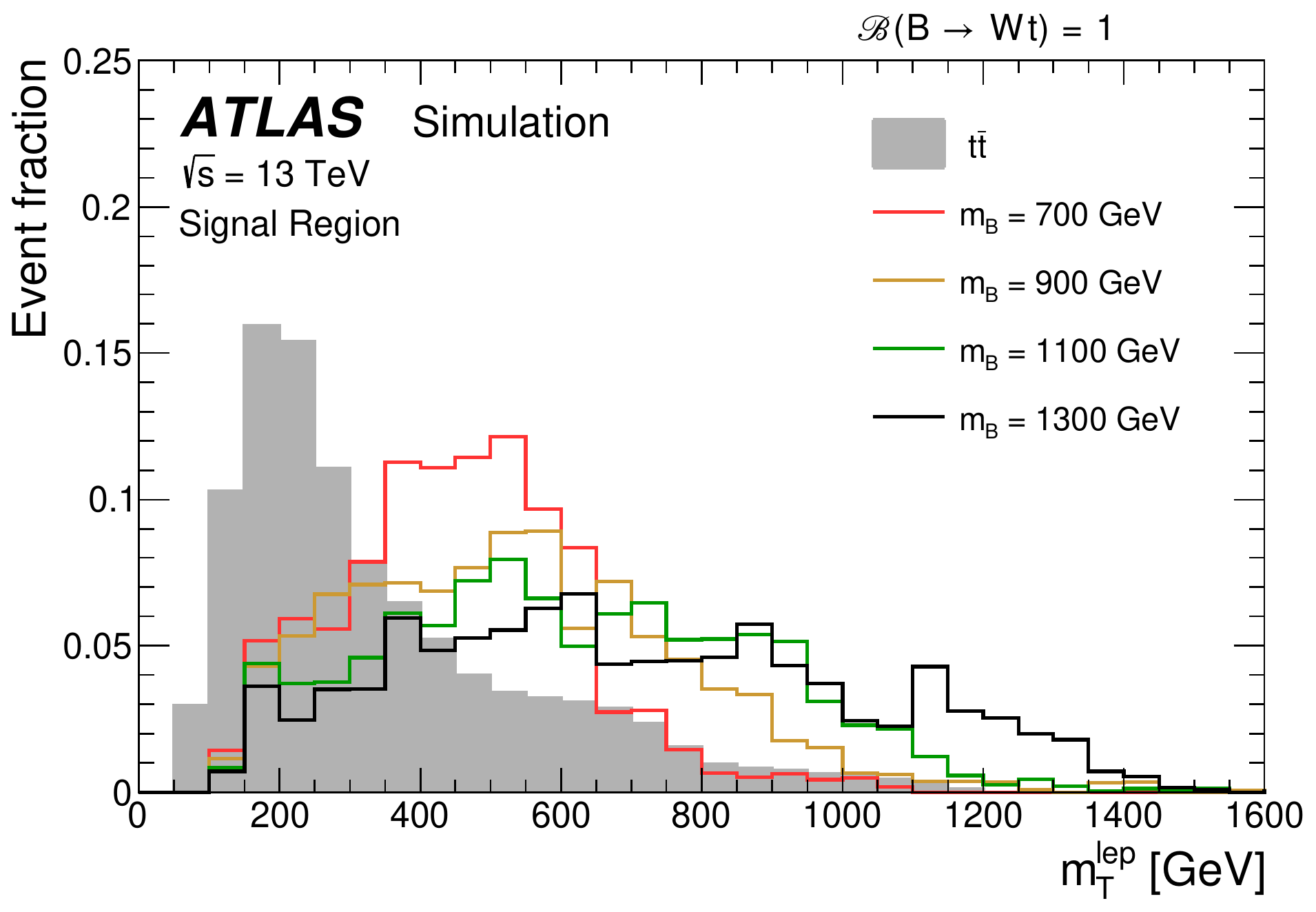}}
\end{minipage}
\caption[]{The reconstructed leptonic VL $T$ mass in the signal region of the $Wb+X$ 
analysis~\cite{wbx} is 
shown for the \ttbar background and a few signal mass points, for signal models assuming $\brtwb=1$ 
(left) and for signal models assuming $\brbwt=1$ (right). In both figures, the 
distributions are normalized to unity for comparison of the relative shapes at each mass point.}
\label{fig:WbX:mTlep}
\end{figure}

The final discriminant used in the analysis is the reconstructed mass of the semi-leptonically 
decaying VL $T$ (\mtlep). The analysis attempts to reconstruct two $T$ candidates in each event by 
making all combinations of the hadronic and leptonic $W$ candidates with the jets in the event 
and selecting the combination that minimizes the quantity $|\mthad-\mtlep|$, where \mthad is the 
reconstructed mass of the fully hadronically decaying VL $T$. Shown 
in the left of Figure~\ref{fig:WbX:mTlep} are 
the unit-normalized \mtlep distributions for the dominant $\ttbar$ background and for signal, 
generated under various assumptions of VL $T$ mass for a model assuming $\brtwb=1$. As can be seen, 
\mtlep provides an excellent separation power against background. The same procedure can be applied 
to a search for VL $B\rightarrow Wt$ without need for further optimization. As shown on the right 
of Figure~\ref{fig:WbX:mTlep} the \mtlep still has a very good separation power against the 
background when used for signal with a model assuming $\brbwt=1$.

\begin{figure}
\begin{minipage}{0.5\linewidth}
\centerline{\includegraphics[width=0.7\linewidth]{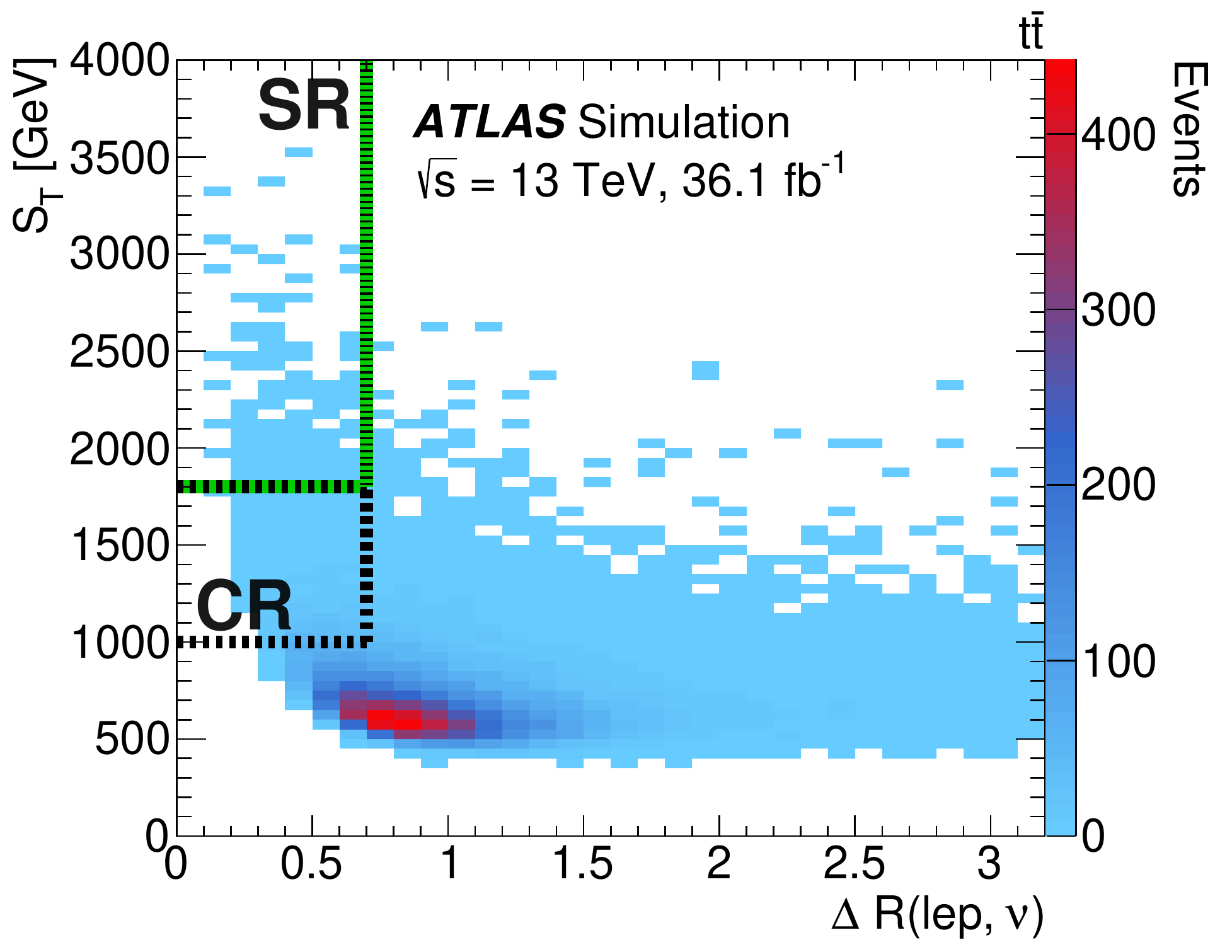}}
\end{minipage}
\hfill
\begin{minipage}{0.5\linewidth}
\centerline{\includegraphics[width=0.7\linewidth]{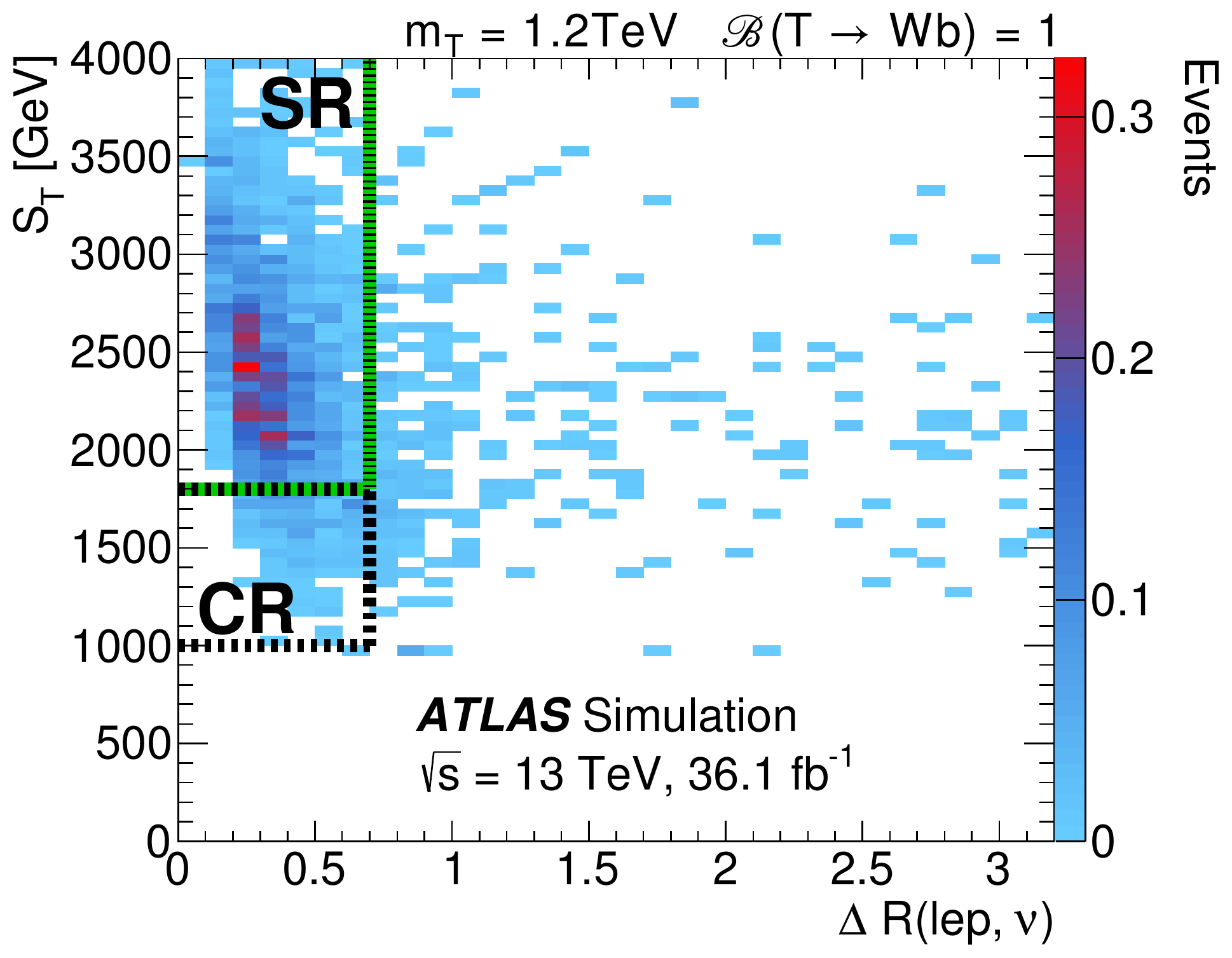}}
\end{minipage}
\caption[]{The signal region (SR) and control region (CR) for the  $Wb+X$  analysis~\cite{wbx} are 
shown in a 
two-dimensional plane of $S_\mathrm{T}$ and $\Delta R(\mathrm{lep}, \nu)$, overlaying 
the distribution of the dominant $t\bar{t}$ background (left) and overlaying the expected 
signal distribution for $\mathcal{B}(T \rightarrow W b)= 100\%$ and a mass of 1.2~TeV (right). }
\label{fig:WbX:srcr}
\end{figure}

The \ttbar background is constrained by a dedicated control region (CR) and the search is performed 
in an orthogonal signal region (SR), by performing a likelihood fit of the background and signal 
contributions to the \mtlep distribution observed in data. The SR and CR are defined using the 
scalar sum of the \met and the \pt of the lepton and jets ($S_\mathrm{T}$) and the separation 
between the lepton and the neutrino ($\Delta R (lep, \nu)$). The definitions of the regions on the 
two-dimensional $\Delta R (lep, \nu)$ -- $S_\mathrm{T}$ plane are shown in 
Figure~\ref{fig:WbX:srcr} which shows the distribution expected for simulated \ttbar background 
(left) and a signal model assuming $\brtwb=1$ and \mbox{$m_T=1.2~\mathrm{TeV}$} (right). As can be 
seen, 
the CR and SR definitions allow for high signal efficiency in the SR and a large number of 
background events in the CR while being as close as possible to the SR. Sub-dominant backgrounds in 
this analysis include multijet events, estimated with the matrix method technique~\cite{mm}, and  
other SM backgrounds 
($W$+jets, single $t$, $Z$+jets, \ttbarV) which are estimated with Monte Carlo.

\begin{figure}
\begin{minipage}{0.5\linewidth}
  \centerline{\includegraphics[width=0.6\linewidth]{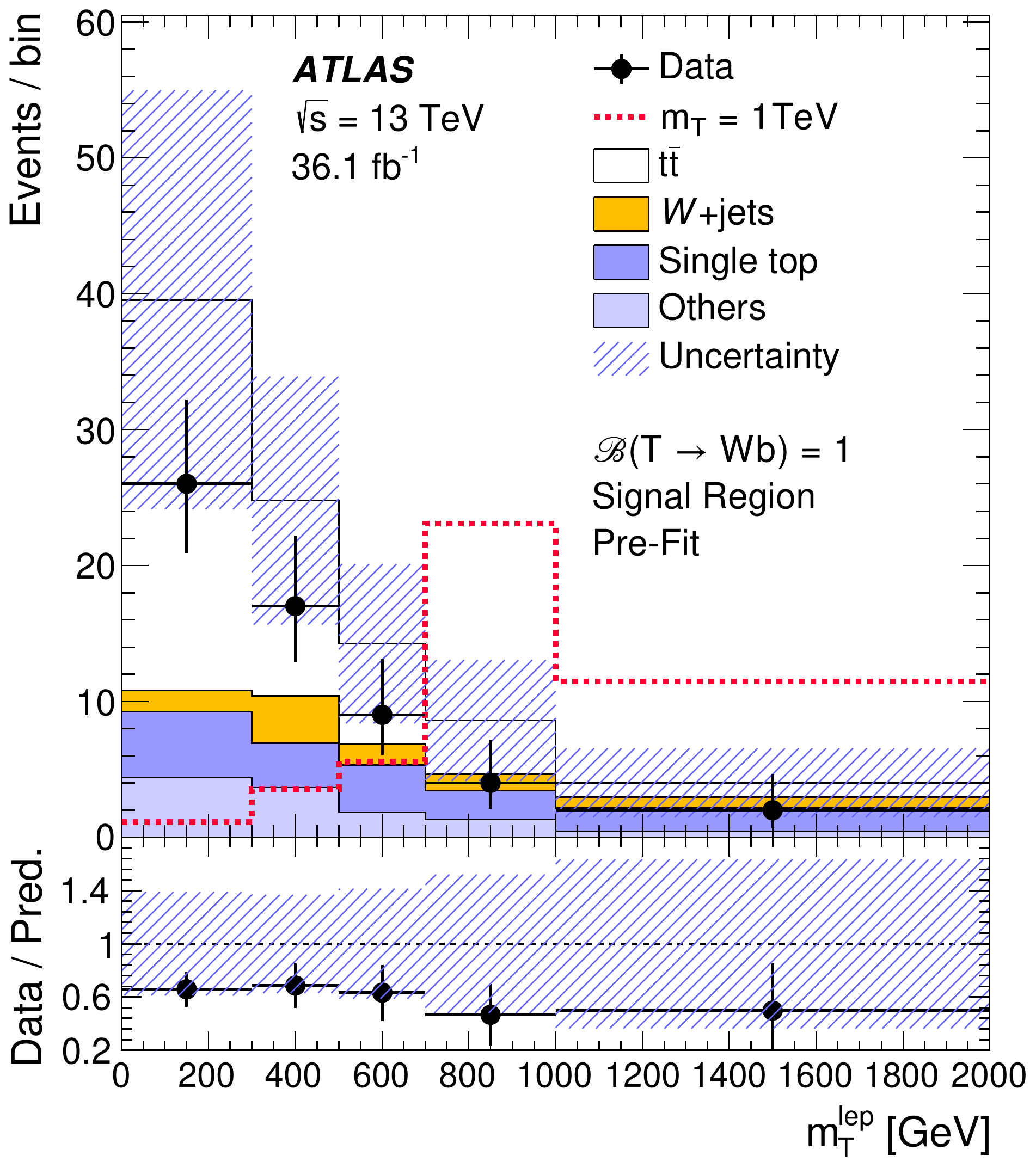}}
\end{minipage}
\hfill
\begin{minipage}{0.5\linewidth}
\centerline{\includegraphics[width=0.6\linewidth]{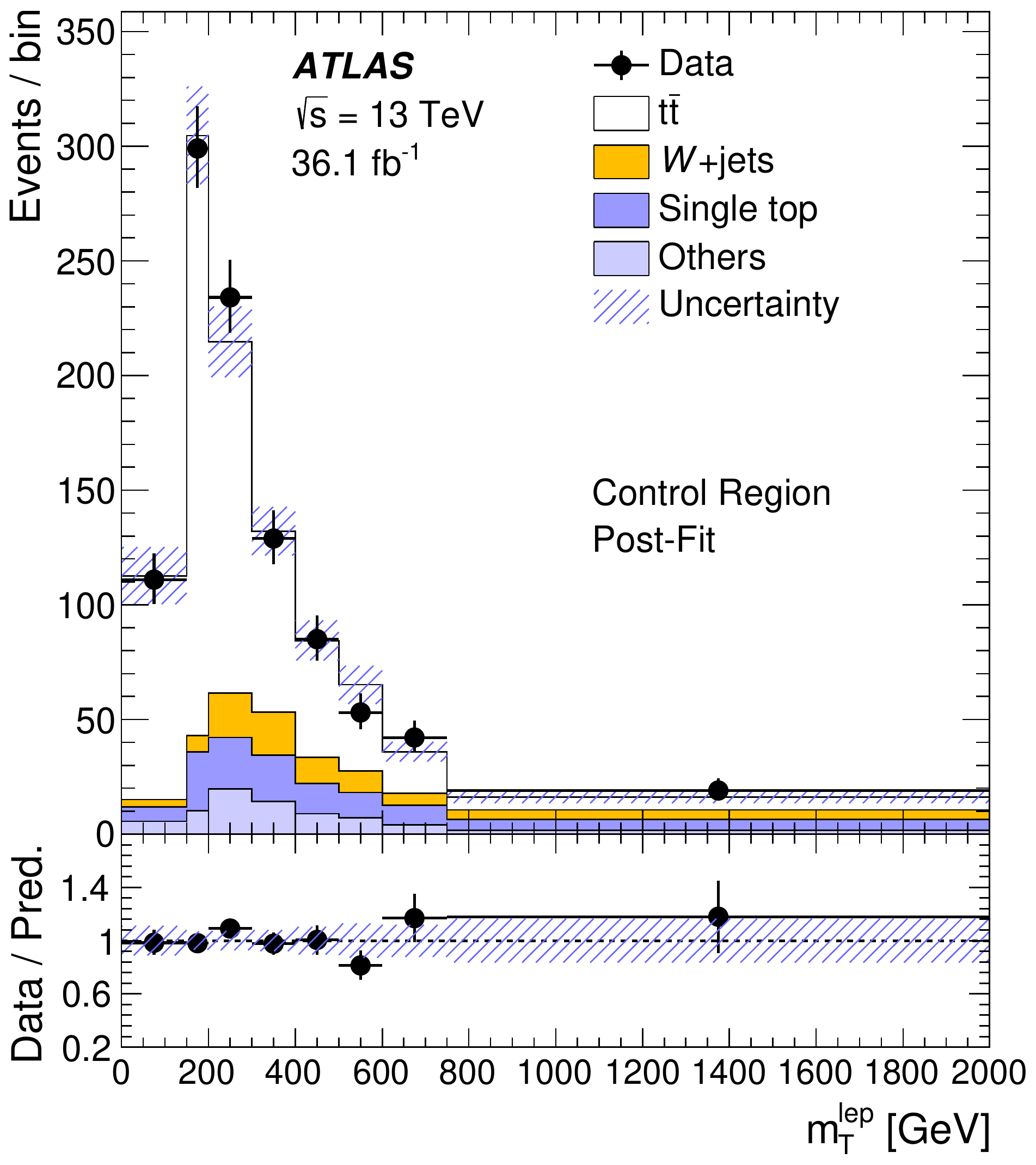}}
\end{minipage}
\caption[]{Leptonic VLQ candidate mass distributions ($m_T^\mathrm{lep}$) 
in the signal region of the  $Wb+X$  analysis~\cite{wbx} before (left) and after a fit under the 
background-only hypothesis (right). The hatched area represents the total uncertainty in the 
background.  }
\label{fig:WbX:prepostfit}
\end{figure}

The dominant systematic uncertainties in this analysis are the single $t$ and \ttbar modelling 
uncertainties and the jet energy resolution. The \mtlep distribution in the SR is shown in 
Figure~\ref{fig:WbX:prepostfit} for background and for data before (left) and after (right) a 
background only fit. The expected signal distribution for a model with $\brtwb=1$ and $m_T=1~TeV$ 
is also shown in the pre-fit plot. After the fit, a very good agreement of the background 
expectation with the data is observed and therefore limits are set at 95\% CL.

\begin{figure}
\begin{minipage}{0.5\linewidth}
\centerline{\includegraphics[width=0.7\linewidth]{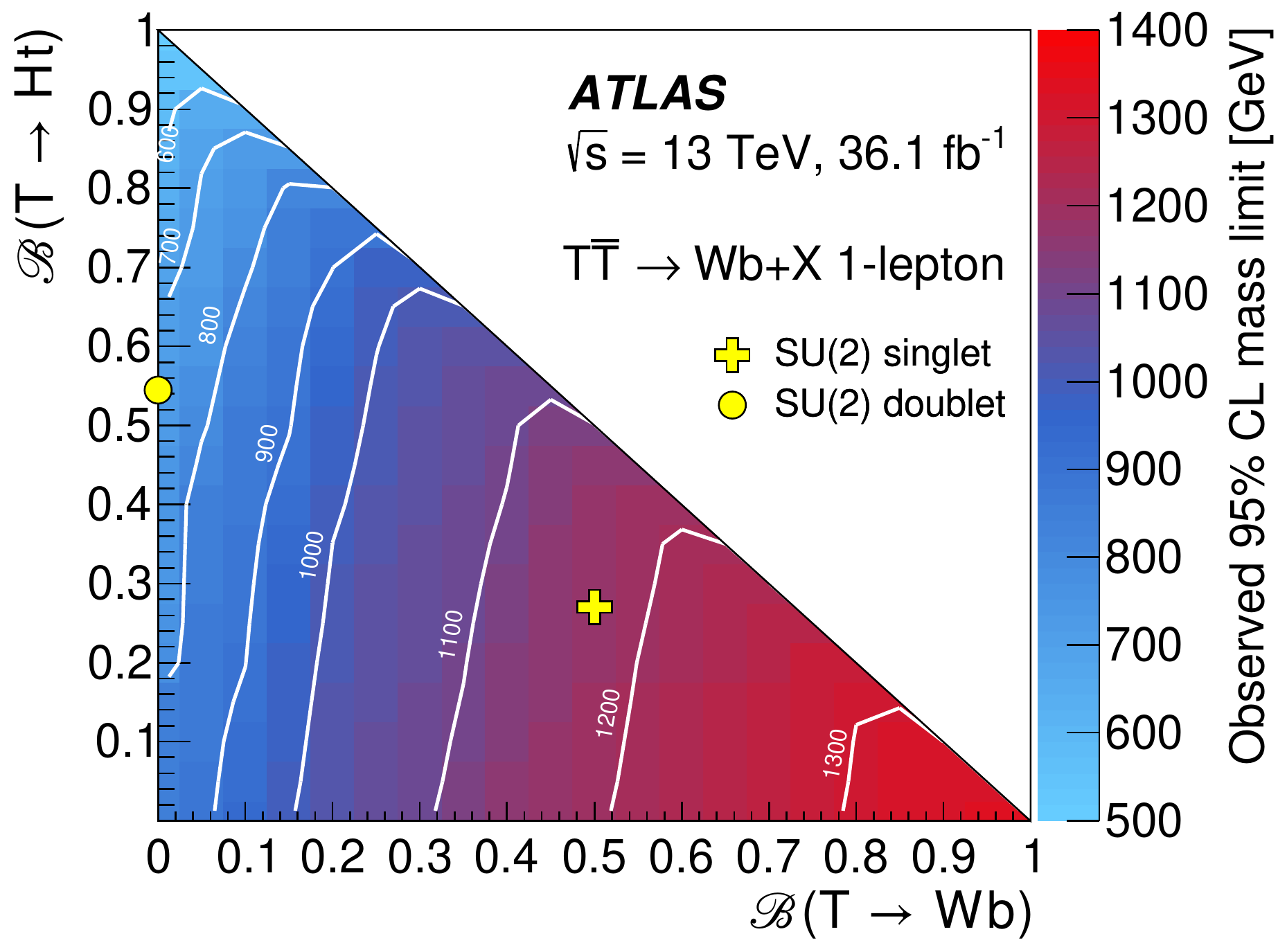}}
\end{minipage}
\hfill
\begin{minipage}{0.5\linewidth}
\centerline{\includegraphics[width=0.7\linewidth]{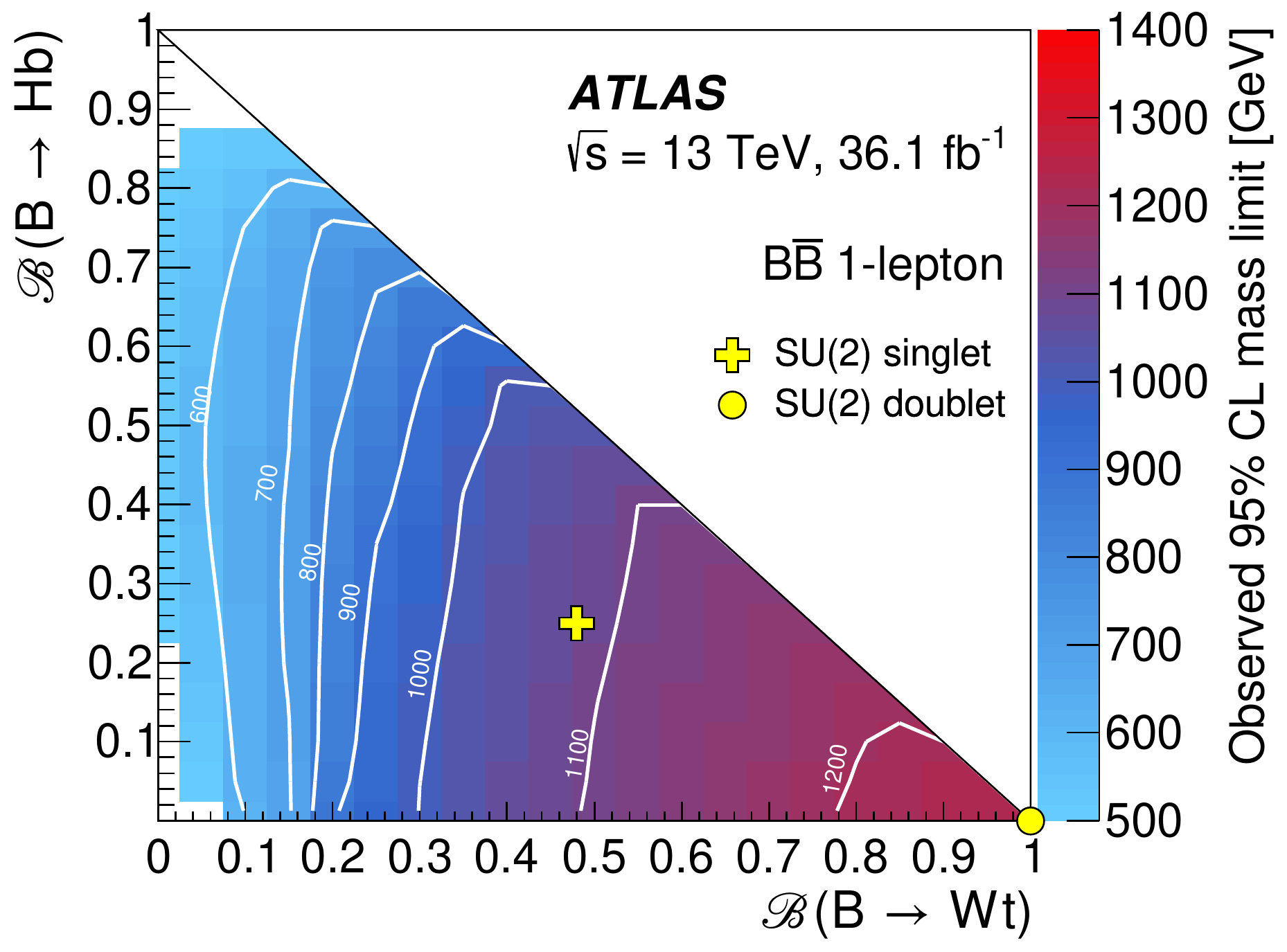}}
\end{minipage}
\caption[]{Observed 95\% CL lower limits on the mass of the $T$ quark in 
the branching-ratio plane of \brtwb versus \brtht 
(left) and observed 95\% CL lower limits on the mass of the $B$ quark in the 
branching-ratio plane of \brbwt versus \brbhb 
(right)  for the $Wb+X$ analysis~\cite{wbx}. The markers indicate the branching ratios for the 
SU(2) singlet and doublet scenarios with masses above $\approx0.8$~TeV, where they are 
approximately independent of the VL $T$ and $B$ masses. The white region is due to the 
limit falling below 500~GeV, the lowest simulated signal mass. 
 }
\label{fig:WbX:brscan}
\end{figure}

As in the first analysis, limits are set for particular benchmark models with specific BR 
assumptions as well as in the two-dimensional BR plane, using a similar signal reweighting 
procedure. Figure~\ref{fig:WbX:brscan} (left) shows the highest excluded VL $T$ mass for a given BR 
composition. The excluded masses are higher near point $(1,0)$ indicating higher sensitivity to 
high \mbox{\brtwb}. Masses up to 1350~GeV are excluded assuming $\brtwb=1$. For the SU(2) singlet 
scenario
with BR to $Wb$, $Zt$, and $Ht$ approximately equal to 0.5, 0.25, and 0.25, respectively, VL 
$T$ masses up to 1170~GeV are excluded. The $\brtwb=1$ limits can be further applied to the VL $Y$ 
quark with charge $q=-4/3e$ which decays exclusively to $Wb$, since no assumption on the VLQ charge 
is made. Pair 
production of VL $Y$ quarks is therefore excluded up to 1350~GeV.

Furthermore, as discussed above, the same analysis can be used to determine limits on the pair 
production of VL $B$ quarks and limits on the two-dimensional \brbwt--\brbhb plane are shown in 
Figure~\ref{fig:WbX:brscan} (right). Assuming $\brbwt=1$, VL $B$ masses are excluded up to 
1250~GeV, while  masses up to 1080~GeV are excluded for a singlet scenario. The former limits are 
also applicable to the VL $X$ with charge $q=5/3e$ which decays exclusively to $Wt$ and is 
therefore 
excluded for masses up to 1250~GeV.

\section{Summary and conclusions}

The latest ATLAS searches for new heavy quarks have focused on VLQs with a broad 
program targeting both pair and single production. Two recent searches for 
the pair production of VLQs performed by ATLAS and targeting decay modes via the $W$, $Z$, and 
$H$ bosons have been presented. Assuming 100\% BR to these decay modes, VL $T$ masses up to between 
1.17~TeV and 1.43~TeV are excluded at 95\% CL. Additionally, VL $B$ masses are excluded up to 
1.25~TeV assuming 100\% BR to $Wt$. Finally, exclusion limits are set for scenarios with 
intermediate BR 
compositions and interpretations in the two-dimensional BR plane for the highest excluded VLQ 
masses are provided.

\let\thefootnote\relax\footnote{Copyright 2018 CERN for the benefit of the ATLAS Collaboration. 
CC-BY-4.0 license.}

\end{document}